# *A full vectorial mapping of nanophotonic light fields*

B. le Feber[1,2], J. E. Sipe[3], M. Wulf[2,4], L. Kuipers[2,5] & N. Rotenberg[2,6]

[1]Optical Materials Engineering Laboratory, ETH Zürich, 8092 Zurich, Switzerland

[2]Center for Nanophotonics, AMOLF, Science Park 104, 1098 XG Amsterdam, The Netherlands

[3]Institute for Optical Sciences, University of Toronto, 60 St. George Street, Ontario M5S 1A7, Canada

4Institute of Science and Technology, Austria, 3400 Klosterneuburg, Austria

[5]Kavli Institute of Nanoscience, Department of Quantum Nanoscience, Delft University of Technology, Lorentzweg 1, 2628 CJ Delft, The Netherlands

[6]Niels Bohr Institute, University of Copenhagen, Blegdamsvej 17, DK-2100 Copenhagen, Denmark

**Light is a union of electric and magnetic fields, and nowhere is their complex relationship more evident than in the near fields of nanophotonic structures. There, complicated electric and magnetic fields varying over subwavelength scales are generally present, leading to photonic phenomena such as extraordinary optical momentum, super-chiral fields, and a complex spatial evolution of optical singularities. An understanding of such phenomena requires nanoscale measurements of the complete optical field vector. However, while it was recently demonstrated that near-field scanning optical microscopy is sensitive to the complete electromagnetic field, a separation of the different components required *a priori* knowledge of the sample. Here we introduce a robust algorithm that can disentangle all six electric and magnetic field components from a single near-field measurement, without any numerical modeling of the structure. As examples, we unravel the fields of two prototypical nanophotonic structures: a photonic crystal waveguide and a plasmonic nanowire. These results pave the way to new studies of complex photonic phenomena at the nanoscale, and for the design of structures that optimize the optical behavior that they exhibit.**

The advent of metamaterials and structures with a large response to the optical magnetic field ushered in a new age of near-field microscopy, one where the ability to measure only electric near fields is no longer sufficient. Many nanoscopic structures, such as split ring resonators [1, 2], dielectric Mie scatterers [3-5], or even simple plasmonic holes [6, 7], have an optical response that depends on the full electromagnetic field. Likewise, measurements of many nanoscale photonic phenomena, such as super-chiral fields [8, 9] and extraordinary spin or orbital angular momentum [10-12], require access to both the electric $\mathbf{E}$ and magnetic $\mathbf{H}$ fields.

Motivated by this need, there have been a number of efforts to extend the capability of near-field scanning optical microscopes (NSOMs) beyond the traditional measurements of $\mathbf{E}$ [13]. Proof-of-concept measurements of $\mathbf{H}$ at the nanoscale have relied on specially designed near-field probes [14, 15]; however, these are difficult to fabricate and tend to measure only one component of $\mathbf{H}$. Recent strategies have therefore focused on measurements with traditional aperture probes [16, 17], culminating in a demonstration that even circular apertures are simultaneously sensitive to the four in-plane components, $E_{x,y}$ and $H_{x,y}$ [18].

Yet a crucial challenge remains. Although a polarization-resolved NSOM measurement (see Supplementary Note 1) contains information from the four in-plane components, it is encoded into only two complex signals $L_x$ and $L_y$, as shown in Fig. 1. To date, unraveling these measurements to extract the individual components of the electric and magnetic fields has not been possible without the use of additional information coming from detailed simulations of the structure being measured [19], or on a symmetry plane where one of the components is identically zero [20]. At best, numerical simulations could be used to determine the spatial evolution of $|\mathbf{E}|^2$ and $|\mathbf{H}|^2$, but not the separate electromagnetic components or their phases [21]. Here we show how to simultaneously extract $E_x$, $E_y$, $H_x$, and $H_y$ from a single, two-channel, NSOM measurement, without any *a priori* knowledge of the nanophotonic structures being measured. By inserting these fields into Maxwell's equations we can obtain the two out-of-plane components $E_z$ and $H_z$, and thus achieve a full vectorial measurement of the electromagnetic near-field. The separation algorithm is robust to noise and realistic measurement conditions, as we show from exemplary NSOM experiments on both photonic crystal waveguides and plasmonic nanowires.

At the heart of near-field microscopy lies the process by which the near-field probe images light fields above a structure. Consider, for example, the field distributions shown in Fig. 1b, which were measured 280 nm above a photonic crystal waveguide (Supplementary Notes 1 and 2), a representative height where the electric and magnetic field distributions are expected to differ [18]. These images are produced as the aperture probe, which acts as an effective spatial filter, merges all four in-plane components of the sample's near field. When this light field is highly structured, with feature sizes smaller than the probe aperture, this process becomes increasingly complex, and it is less obvious is exactly how efficiently and with what phase $E_x$, $E_y$, $H_x$, and $H_y$ contribute to the measured signals $L_y$ and $L_x$. That is, calculating the transfer function of a near-field probe, which would propagate the fields from the sample to a detector, has not been possible.

What is possible, however, is to calculate the fields radiated through the probe by a point dipole placed at the position $\mathbf{r}_0$ of a hypothetical detector (Fig. 2a), with current density $\mathbf{j}_{det}\delta(\mathbf{r}-\mathbf{r}_0)$. These fields, which we label $\mathbf{E}_x^r$ and $\mathbf{H}_x^r$ (Fig. 2b, middle column, for the dipole lying in the $x$ direction) have been measured extensively, and resemble those below a hole in a metal film [22, 23]. Then, via the optical reciprocity theorem (ORT), we can use these probe fields to relate the sample fields $\mathbf{E}^e$ and $\mathbf{H}^e$ (Fig. 2b, left column) to dipoles that would be induced at our detectors, and hence to our measured signals (Fig. 2b, right column) [18, 24-26]. That is, in this approach $\mathbf{E}_i^r$ and $\mathbf{H}_i^r$, with $i = x, y$ indicating the orientation of $\mathbf{j}_{det}$, can be viewed as the spatial filters that define exactly how efficiently and with what phase the different sample field components are detected. The fact that each independent dipole orientation, $x$ or $y$, is associated with all four in-plane components of the probe field explains why each detection channel typically contains information of all the in-plane components of the sample fields. It is possible, using a specific sensing configuration [17] or material composition [26], to design probes that primarily detect $\mathbf{E}^e$ or $\mathbf{H}^e$ of specific near fields. Such probes, however, preclude complete electromagnetic measurements and therefore, in this work, we consider aperture probes that are similarly sensitive to $\mathbf{E}^e$ and $\mathbf{H}^e$.

Image formation via the ORT can be expressed as (see Supplementary Note 4 for derivation)

$$L_i\left(\mathbf{R}_{tip}\right) = \int_S dS \left(\mathbf{E}^e(\mathbf{R}) \times \mathbf{H}_i^r\left(\mathbf{R}-\mathbf{R}_{tip}\right) - \mathbf{E}_i^r\left(\mathbf{R}-\mathbf{R}_{tip}\right) \times \mathbf{H}^e(\mathbf{R})\right) \cdot \hat{\mathbf{z}}, \qquad (1)$$

where $S$ is a surface between the probe and the sample, which is here 10 nm below the probe, $\mathbf{R}_{tip} = (x_{tip}, y_{tip})$ is the position of the tip above this plane, $\mathbf{R} = (x, y)$ are the coordinates of the fields on $S$ and the integral is taken over all $\mathbf{R}$. Note that subscript $i$ refers to the $x$ or $y$ orientation of the reciprocal dipole, not a component of the fields. The dot product with $\hat{\mathbf{z}}$ shows that the measured image depends only on the in-plane field components. This process of image formation is shown in Fig. 2b, where we use calculated probe and sample fields to predict the measured signals. In fact, we see excellent agreement between our predictions (right column, Fig. 2b) and the measurements (Fig. 1b) 280 nm above the PhCW.

When we want to retrieve the sample fields, rather than study the image formation, we face two challenges: First, we require two additional equations to match the number of unknowns; and second, we must be able to invert Eq. 1 (Supplementary Note 3). To deal with the first challenge we recognize that the electromagnetic field at and near the sample plane can be decomposed into a superposition of different plane waves, each represented by a total wavevector $\mathbf{k} = k_z \hat{\mathbf{z}} + \kappa \hat{\boldsymbol{\kappa}}$ [27]. Here $k_z$ is the out-of-plane component of the wavevector, and $\boldsymbol{\kappa}$ the in-plane component, as shown in Fig. 2a. We can write each plane wave in the Cartesian basis ($E_x$, $E_y$, $E_z$) or in terms of its *s*- and *p*-field components ($E_{s+}$, $E_{s-}$, $E_{p+}$, $E_{p-}$), which allows us to identify the upwards ($\text{real}(k_z) > 0$, subscript $+$) or downwards ($\text{real}(k_z) < 0$, subscript $-$) propagating waves. Since above the sample surface only upwards propagating fields exist (i.e. $E_{p-} = 0$), we need only consider the first two components of the electric field, and hence the total field can be written in terms of only $E_s$, $E_p$, $H_s$, and $H_p$, where all $s$ and $p$ components are understood to be upwards propagating (i.e. $p+$). Finally, Maxwell's equations straightforwardly relate the electric and magnetic field components of these transverse plane waves (see Supplementary Note 5 for derivation and conversion between the different bases)

$$\begin{aligned} \mathbf{E}^e(\boldsymbol{\kappa}) &= E_s^e(\boldsymbol{\kappa})\hat{\mathbf{s}} + E_p^e(\boldsymbol{\kappa})\hat{\mathbf{p}}, \\ \mathbf{H}^e(\boldsymbol{\kappa}) &= \frac{1}{Z_0}\left[ E_s^e(\boldsymbol{\kappa})\hat{\mathbf{s}} - E_p^e(\boldsymbol{\kappa})\hat{\mathbf{p}} \right], \end{aligned} \quad (2)$$

where $Z_0$ is the impedance of free space. In light of Eq. 1, we have now reduced our problem to two unknowns $E_s^e$ and $E_p^e$ and two equations, one each for $L_x$ and $L_y$. In terms of the Fourier components we can then rewrite Eq. 1 as

$$\frac{1}{Z_0}\begin{bmatrix} L_x^{\text{det}}(\boldsymbol{\kappa}) \\ L_y^{\text{det}}(\boldsymbol{\kappa}) \end{bmatrix} = \begin{bmatrix} N_{x,s}(\boldsymbol{\kappa}) & N_{x,p}(\boldsymbol{\kappa}) \\ N_{y,s}(\boldsymbol{\kappa}) & N_{y,p}(\boldsymbol{\kappa}) \end{bmatrix} \begin{bmatrix} E_s^e(\boldsymbol{\kappa}) \\ E_p^e(\boldsymbol{\kappa}) \end{bmatrix}, \quad (3)$$

where the tensor $\mathbf{N}$ is essentially the transfer matrix that maps the sample electric fields expressed in their polarization components to the detection channels associated with the $x$- and $y$-directions. The different components of $\mathbf{N}$ are related to the *Cartesian* components of $\mathbf{E}_i^r$ and $\mathbf{H}_i^r$ by

$$\begin{bmatrix} N_{i,s}(\boldsymbol{\kappa}) \\ N_{i,p}(\boldsymbol{\kappa}) \end{bmatrix} = \begin{bmatrix} -\frac{k_z}{k_0}\sin\varphi & \frac{k_z}{k_0}\cos\varphi & Z_0\cos\varphi & Z_0\sin\varphi \\ \cos\varphi & \sin\varphi & Z_0\frac{k_z}{k_0}\sin\varphi & -Z_0\frac{k_z}{k_0}\cos\varphi \end{bmatrix} \begin{bmatrix} E_{i,x}^r(-\boldsymbol{\kappa}) \\ E_{i,y}^r(-\boldsymbol{\kappa}) \\ H_{i,x}^r(-\boldsymbol{\kappa}) \\ H_{i,y}^r(-\boldsymbol{\kappa}) \end{bmatrix}, \quad (4)$$

where $\varphi$ is the angle $\boldsymbol{\kappa}$ makes from the $x$-axis (Fig. 2a). We show the process of image formation in terms of these plane wave components in the top row of Fig. 2c — corresponding to the real space plots in Fig. 2b — where $N_{x,s}(\boldsymbol{\kappa})$ and $N_{x,p}(\boldsymbol{\kappa})$ are plotted in the middle column. From these $\mathbf{N}$ maps, it is clear which wavevector components contribute most to the detected image.

Unraveling the near-field measurements is then simply a matter of inverting $\mathbf{N}$ to obtain

$$\begin{bmatrix} E_s^e(\boldsymbol{\kappa}) \\ E_p^e(\boldsymbol{\kappa}) \end{bmatrix} = \frac{1}{Z_0}\begin{bmatrix} N_{x,s}(\boldsymbol{\kappa}) & N_{x,p}(\boldsymbol{\kappa}) \\ N_{y,s}(\boldsymbol{\kappa}) & N_{y,p}(\boldsymbol{\kappa}) \end{bmatrix}^{-1} \begin{bmatrix} L_x^{\text{det}}(\boldsymbol{\kappa}) \\ L_y^{\text{det}}(\boldsymbol{\kappa}) \end{bmatrix}, \quad (5)$$

which has a unique solution if $\det(\mathbf{N}) \neq 0$, for all $\boldsymbol{\kappa}$, as is indeed the case for our probes. We can therefore deconvolve a near-field measurement simply by following the steps illustrated in the bottom row of Fig. 2c. First, the measurements are Fourier transformed in the *xy*-plane, generating $L_{x,y}(\boldsymbol{\kappa})$. These are multiplied by $\mathbf{N}^{-1}(\boldsymbol{\kappa})$ to obtain $E_{s,p}^e(\boldsymbol{\kappa})$ according to Eq. 5. These fields are then transformed back into the Cartesian basis (Supplementary Note 5), and inverse Fourier transformed back into real space, to arrive at the deconvolved sample fields $E_{x,y}^e(\mathbf{R})$ and $H_{x,y}^e(\mathbf{R})$. Finally, following the example of Olmon et al. [20] we make use of Maxwell's equations to extract the 2D maps of the out-of-plane electric and magnetic field components, $E_z^e(\mathbf{R})$ and $H_z^e(\mathbf{R})$, according to $E_z = iZ_0k_0\left(\frac{\partial H_x}{\partial x} - \frac{\partial H_x}{\partial y}\right)$ and $H_z = -\frac{ik_0}{Z_0}\left(\frac{\partial E_y}{\partial x} - \frac{\partial E_x}{\partial y}\right)$. Because the same probe can be used for multiple measurements, and since $\mathrm{N}(\boldsymbol{\kappa})$ is similar for probes with differing aperture sizes (Supplementary Figure S12), $\mathrm{N}^{-1}(\boldsymbol{\kappa})$ needs only to be calculated once and can then be used in many experiments.

Note that the inversion of $\mathrm{N}$ (in Eq. 5) makes our deconvolution process sensitive to large-wavevector signals, even though the image formation process is not (bottom and top rows of Fig. 2c, respectively). Since the experimental fields (left column, Fig. 2c) do not contain signal at these large wavevectors, it is there that measurement noise typically dominates. While in principle this sensitivity to large wavevectors limits our retrieval algorithm, in practice it does not greatly affect its performance. As we discuss below (see Fig. 4), we can simply limit the largest wavevector that we consider to that wavevector at which we still expect to find signal from the sample.

Hence, when we apply our algorithm to the PhCW fields shown in Fig. 1b, we limit $\mathrm{N}^{-1}$ to the region where $\kappa \leq 3k_0$, where $k_0$ is the free space wavenumber of the light. The amplitudes of the separated field components are shown in Fig. 3 along with the theoretically calculated mode profiles. Line cuts, taken at the positions of the dashed lines, are also shown, demonstrating the excellent agreement when comparing the experimental (blue) and theoretical (grey) curves for all six electromagnetic field components. In fact, the only component for which we observe significant deviation between the predicted and measured field amplitude is $E_z$. We attribute this difference to the small amplitude of this component, which makes it more susceptible to errors that arise from imperfect experimental conditions that could lead to, e.g., polarization mixing. We also observe strikingly good agreement between the calculated and retrieved phase profiles (Supplementary Figure S14). That is, not only are we able to successfully recover the general shape of each field component, but we can even resolve the fine features in the amplitude and phase of these in-plane fields, all from a *single* measurement.

Our approach is not limited to dielectric structures, but can be extended to the realm of nanoplasmonics. As an example we consider a plasmonic nanowire, whose electric and magnetic near-field distributions are known to have different, and none-trivial, spatial dependencies [28]. Using our protocol we resolve the different field components above the nanowire (see Supplementary Note 6 for details and images of the separated fields). We again observe good agreement between the theoretical and measured fields and, as we found for dielectric samples, clear differences in the retrieved electric and magnetic fields from different samples are revealed (Supplementary Figure S10).

The ability of our algorithm to retrieve optical fields from measurements of both a PhCW and a plasmonic nanowire already hints at its robustness to noise. To further explore the effect of measurement noise, we artificially add white noise to a perfect 'measurement' (i.e. theoretically calculated fields with a noise level < $10^{-3}$) in increments until we reach a signal-to-noise ratio of unity in $L_{x,y}$. We then calculate the normalized error between the ideal and retrieved optical fields (see Methods Section), which is shown in Fig. 4. More importantly, for all noise levels we observe that setting $\kappa_{max} \leq 2k_0$ results in a poor field retrieval, as this low limit effectively filters large portions of the input signal (Supplementary Figure S12 for corresponding retrieved field maps, and Supplementary Section S7 for additional discussion). However, up to $\kappa_{max} \leq 5k_0$ we find near-perfect deconvolution even in cases where the noise is as large as the signal.

Finally, we note that while decreasing the probe aperture size results in a decrease in signal and a corresponding increase in resolution, it has little effect on our algorithm (Supplementary Figure S13); although higher wave vectors appear in $\mathrm{N}(\boldsymbol{\kappa})$ for small probe diameters, at low $\boldsymbol{\kappa}$ $\mathrm{N}(\boldsymbol{\kappa})$ remains nearly identical. Since the algorithm is robust even when the noise level is comparable to the signal (c.f. Fig. 4) even measurements with such low-throughput probes can be deconvolved into their constituent components.

The capability to map both the electric and magnetic near-field components is important for the study and development of nanophotonic structures, particularly if the strategy is simple and robust. Our approach can be used to measure the full electric and magnetic fields near dielectric and plasmonic structures, increasingly necessary in a research landscape of nanoscopic structures with different electric and magnetic responses. As a demonstration, we have presented the full, complex electromagnetic near-field of two nanophotonic waveguides, but we note that our approach can also be applied to other systems such as nano-antennas and cavities. For the latter case, special care must be taken with high quality factor resonators $Q \gtrsim 1000$, where interactions between the near-field probe and the photonic mode cannot be neglected, and in fact can provide an independent measure of the magnetic field [15, 29]. Measurements of nanoscale $\mathbf{E}$ and $\mathbf{H}$ have the potential to drive progress in fields such as plasmonics [30], on-chip photonics [31, 32] and metasurfaces [33], where simulations of realistic structures with unavoidable imperfections are often not available. A further intriguing possibility is the combination of our method with measurements of the emission of a quantum emitter placed on the probe, which map out the local density of optical states [34, 35], and are therefore important to quantum optical applications.

**Acknowledgements**

The authors thank Irina Kabakova and Anouk de Hoogh for help with experiments and fabrication. The authors acknowledge support from the European Research Council (ERC Advanced Grant 340438-CONSTANS). This work is also part of the research program Rubicon with project number 680-50-1513, which is (partly) financed by the Netherlands Organization for Scientific Research (NWO). Finally, part of this work is also funded by the Natural Sciences and Engineering Research Council of Canada.

**Author contributions**

N.R., K.K. and J.S. conceived the study. J.E.S., N.R., B.l.F. and L.K. developed the deconvolution algorithm. B.l.F performed numerical analyses and simulations. N.R., B.l.F, M.W. and L.K. performed experiments. J.E.S., N.R., B.l.F. and L.K. wrote the manuscript. The authors declare no competing financial interests.

**Methods**

*Robustness to noise*

To quantify the robustness to noise of our algorithm we compare the calculated fields to the fields retrieved from a computer-generated field map obtained by applying the reciprocity theorem on calculated fields. To this calculated mapping (such as shown in Fig. 2b) we add a controlled amount of white noise. The mean amplitude of that noise relative to the maximum amplitude of the signal is shown on the $y$-axis of Fig. 4. Next, we apply our algorithm to these noisy calculated mappings and compare the retrieved fields to the calculated fields, to obtain the normalized error $\Delta = \sum_{E_{x,y} H_{x,y}} \int \left| |F_{retr}| - |F_{in}| \right|^2 dr \Big/ \sum_{E_{x,y} H_{x,y}} \int |F_{in}|^2 dr$, where $F$ indicates the electric and magnetic field components of the retrieved (retr.) and input (in) fields.

*SP coordinate transformations*

The orientation of the $sp$-basis vectors is constructed from the in-plane wavevector according to

$$\hat{\mathbf{s}} = \hat{\mathbf{\kappa}} \times \hat{\mathbf{z}},$$

$$\hat{\mathbf{p}}_{\pm} = \frac{\kappa \hat{\mathbf{z}} \mp k_z \hat{\mathbf{\kappa}}}{k_0},$$

where $k_z = \sqrt{k_0^2 - \kappa^2}$. In our experiment only upwards propagating fields exist, and we use the following equations to convert the fields in the $sp$-basis to those in a Cartesian basis,

$$E_x^e(\mathbf{\kappa}) = \sin\phi \, E_s(\mathbf{\kappa}) - \frac{k_z}{k_0} \cos\phi \, E_p(\mathbf{\kappa}),$$

$$E_y^e(\mathbf{\kappa}) = -\cos\phi \, E_s(\mathbf{\kappa}) - \frac{k_z}{k_0} \sin\phi \, E_p(\mathbf{\kappa}),$$

$$H_x^e(\mathbf{\kappa}) = \sin\phi \frac{E_p(\mathbf{\kappa})}{Z_0} + \frac{k_z}{k_0} \cos\phi \frac{E_s(\mathbf{\kappa})}{Z_0} \text{ and}$$

$$H_y^e(\mathbf{\kappa}) = -\cos\phi \frac{E_p(\mathbf{\kappa})}{Z_0} + \frac{k_z}{k_0} \sin\phi \frac{E_s(\mathbf{\kappa})}{Z_0}.$$

These equations are derived in Supplementary Note 5 and can be straightforwardly inverted to find the transformation from a Cartesian to an $sp$-basis.

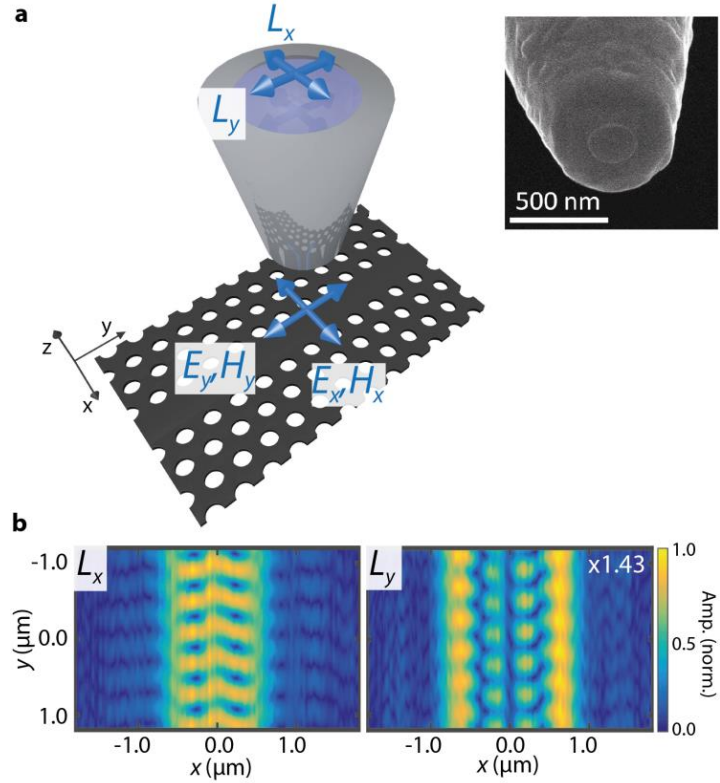

**Figure 1: Polarization-resolved near-field measurements. a** Sketch of the essentials of the polarization-sensitive NSOM used in this work. The blue arrows near the sample indicate electric and magnetic fields along $x$ and $y$. The probe converts these fields to radiation polarized along $x$ and $y$, indicated by the top blue arrows. The inset shows an SEM of the aperture probe used for the photonic crystal waveguide measurements. **b** Two-dimensional maps of the amplitude of $L_x$ (left panel) and $L_y$ (right) measured by raster-scanning the tip 280 nm above the photonic crystal waveguide.

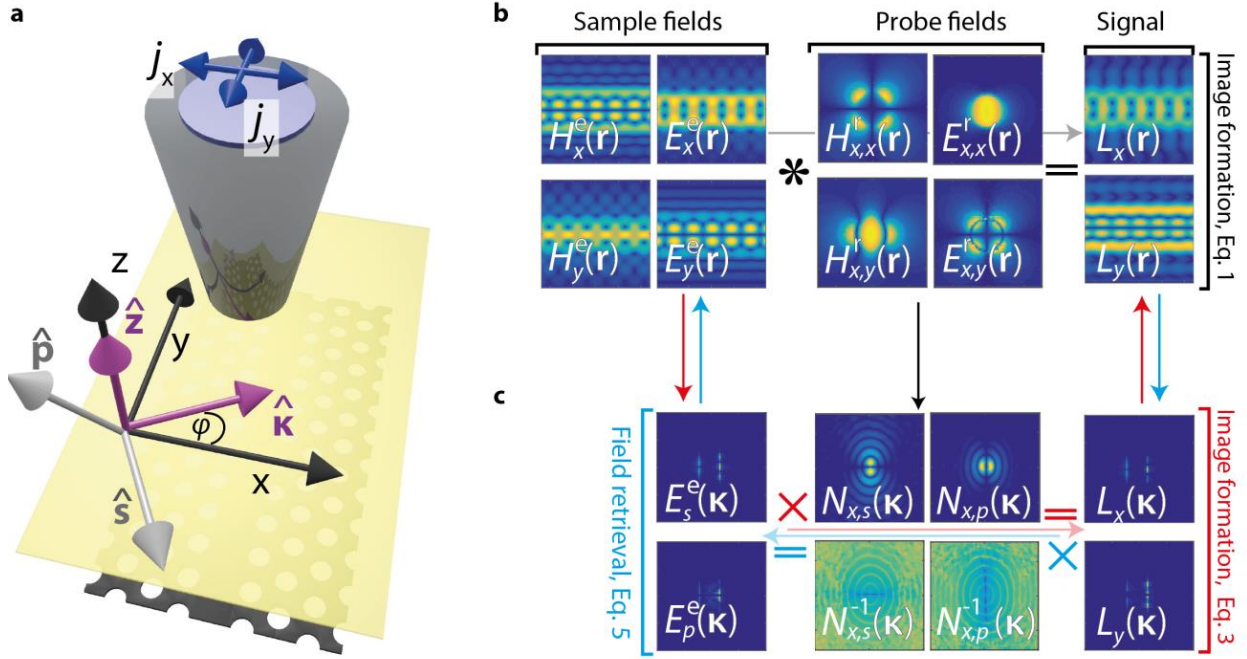

**Figure 2**: **Image formation and field retrieval. a** Schematic of the coordinate bases and experimental setup. All fields are evaluated on a surface (transparent yellow) that completely separates the probe from the sample. The purple arrows indicate the in-plane ($\hat{\boldsymbol{\kappa}}$) and out-of-plane ($\hat{\mathbf{z}}$) unit vectors of a plane wave on this surface, while the gray arrows show the corresponding unit vectors $\hat{\mathbf{s}}$ and $\hat{\mathbf{p}}$ for an upwards travelling wave. **b** Real-space image formation process according to Eq. 1. In real space, the measured image, $L_{x,y}$, can be understood as the convolution (indicated by the $*$ sign) of the sample fields, $\mathbf{E}^e$ and $\mathbf{H}^e$, and the probe fields, $\mathbf{E}_i^r$ and $\mathbf{H}_i^r$, shown here for $x$-oriented dipole ($i=x$). **c** Top row: In Fourier space, the process of image formation corresponding to **b** is described by the multiplication of the sample fields and the probe response function $\mathrm{N}$. Bottom row: The reverse process, which results in the separated fields, therefore simply involves the multiplication of the measured signals with the inverse probe response function $\mathrm{N}^{-1}$. Note that we show only the $x$-oriented dipole ($i=x$) components of $\mathrm{N}$ and $\mathrm{N}^{-1}$. All maps in **b** and **c** show calculated fields that are normalized to their maximum amplitude.

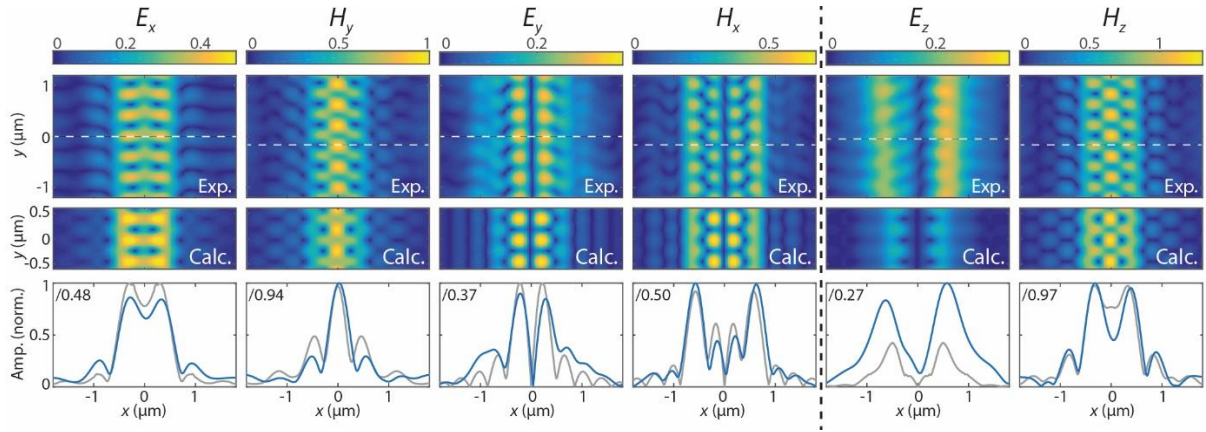

**Figure 3: Retrieved PhCW electric and magnetic fields.** Panels show two-dimensional amplitude maps of the retrieved (top) and calculated (middle) electric and magnetic fields 280 nm above a PhCW. The field components shown in each column are indicated above that column, with the black dashed line separating the in- and out-of-plane fields. The retrieved and calculated amplitudes are normalized to the maximum amplitude of the retrieved $H_y$. In the bottom row of panels we show line cuts taken across the maxima of each field as indicated by the white dashed lines in the field maps. Blue and gray lines correspond to line cuts through the retrieved and calculated fields, respectively. To show all fields on the same axis we scaled the amplitude with the factors show in the top left of each panel.

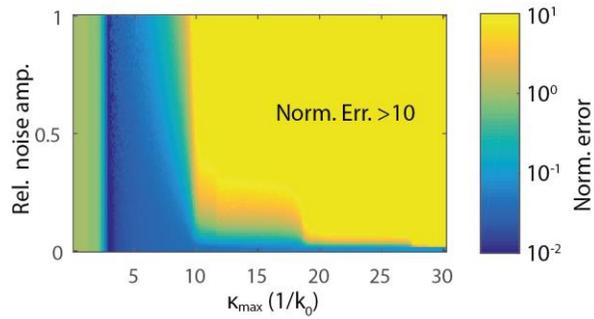

**Figure 4: Robustness of the field retrieval algorithm.** Mismatch between the retrieved fields and the predicted fields (see Methods) as a function of the noise amplitude and wavevector cutoff (see text for explanation). Because small signals with a high spatial frequency can result in very large signals, well beyond the total intensity of the calculated fields, we saturate Fig. 4 at normalized errors larger than 10, to avoid obscuring more important results at low mismatch values. Likewise, the minimum error in our calculations is at $10^{-5}$, practically values below $10^{-2}$ appear identical to the input fields, hence we also saturated this plot below $10^{-2}$.

Supplementary information for:

*A full vectorial mapping of nanophotonic light fields*

Boris le Feber[1,2], John. E. Sipe[3], Matthias Wulf[2,4], L. Kuipers[2,5] & N. Rotenberg[2,6]

[1]Optical Materials Engineering Laboratory, ETH Zurich, 8092 Zurich, Switzerland

[2]Center for Nanophotonics, AMOLF, Science Park 104, 1098 XG Amsterdam, The Netherlands

[3]Institute for Optical Sciences, University of Toronto, 60 St. George Street, Ontario M5S 1A7, Canada

4Institute of Science and Technology Austria, 3400 Klosterneuburg, Austria

[5]Kavli Institute of Nanoscience, Department of Quantum Nanoscience, Delft University of Technology, Lorentzweg 1, 2628 CJ Delft, The Netherlands

[6]Niels Bohr Institute, University of Copenhagen, Blegdamsvej 17, DK-2100 Copenhagen, Denmark

## Supplementary Note 1: Phase- and polarization-resolved NSOM

*Phase-resolved NSOM*

Access to the optical phase can be gained by incorporating the near-field scanning optical microscope (NSOM) in an interferometric detection scheme, as sketched in Fig. S1. Specifically, we use a heterodyne detection scheme, in which the light from the probe interferes with the frequency shifted reference radiation from the reference branch; see Fig. S1. The frequency of the reference branch is shifted by 40 kHz using two acousto-optic modulators (AOMs); see Fig. S1. By analyzing the beating signal measured on the diodes on two lock-in amplifiers, we gain access to the optical phase of the signal branch.

*Polarization-resolved NSOM*

To measure the polarization of the light emitted from the probe tip, we use a polarizing beam splitter, marked by PBS in Fig. S1, which ensures that light polarized along $x$ and $y$ contributes to the signals $L_x$ and $L_y$, respectively. However, to be able to relate light emitted with $x$ and $y$ polarizations near the sample (indicated in Fig. S1) to $x$- and $y$-polarized light (now in the lab frame) at the detectors, we need to correct for birefringence in the fibers after the probe.

As a result of this birefringence, linear $x$- or $y$-polarized radiation from the probe will typically become elliptically polarized upon transmission through the fiber. To project these elliptical polarizations back onto the $x$- and $y$-orientations above the sample, we employ the quarter- and half-wave plates sketched in Fig. S1. First, after the fiber we use the quarter-wave plate ($\lambda/4$) to project the elliptically polarized light back onto linearly polarized light. Then, we insert the second half-wave plate ($\lambda/2_{(2)}$) to rotate the light such that $x$- and $y$-polarized radiation from the probe contributes to $L_x$ and $L_y$, respectively. Finally, we use the first half-wave plate ($\lambda/2_{(1)}$) to balance the intensity of the reference branch over the two detectors.

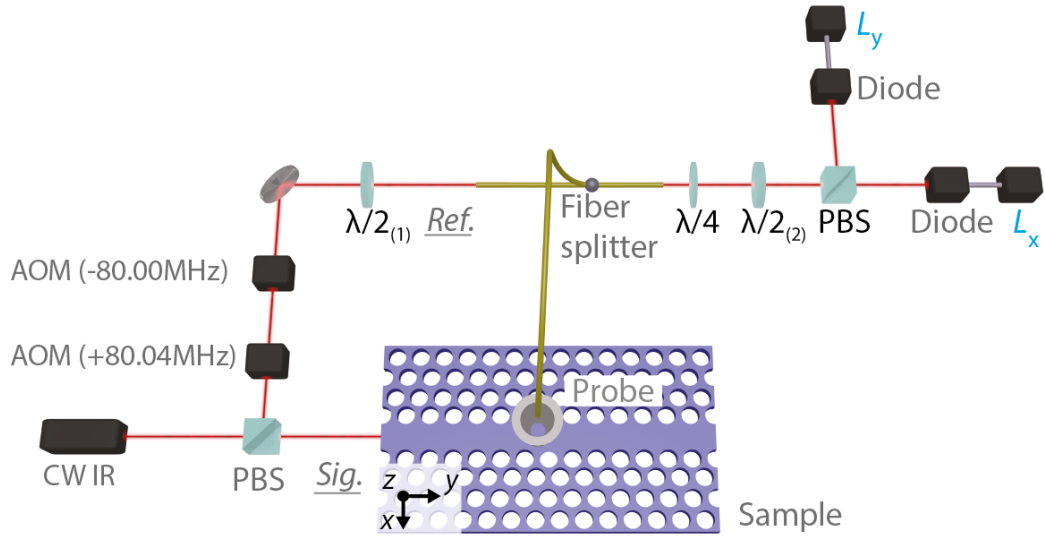

**Supplementary Figure S1. Polarization-sensitive NSOM.** Light from a continuous wave infrared laser (CW IR) is split up into a signal (Sig.) and a reference (Ref.) branch. The reference branch is frequency shifted using two AOMs before it is coupled to a fiber (yellow tube). The light in the signal branch is coupled into the sample, from where the light is collected by a near-field probe. Light from the probe propagates through the fiber, where it joins the reference branch. After the fiber splitter the light from the two branches is converted to a free-space beam, which, after passing through the polarizing beam splitter (PBS), is detected on the photodiodes, whose signal is analyzed by the lock-in detectors. This extension adds polarization sensitivity by means of the elements marked with black letters. The elements required for a phase-sensitive NSOM are marked in gray. In blue we indicate the two signals $L_x$ and $L_y$.

## *Height-feedback mechanism*

We use a force feedback loop to keep the probe at the sample when scanning the surface of the photonic crystal while "in contact". However, if $h > 20$ nm, we can no longer use the force feedback, so we therefore switch to a quadrant-cell-based height-feedback loop. Here, the quadrant-cell that measures the relative position of the probe towards or away from the sample, is fixed to the same frame as the sample and does not move whilst the probe scans the sample. When feeding back on the quadrant cell, we feed back on the distance away from the sample compared to the probe-sample-distance, as measured with the quadrant cell, whilst scanning the surface of the sample [1].

## Supplementary Note 2: Photonic-crystal waveguide mode calculation

The photonic crystal waveguide is a W1 waveguide, which has a row of missing holes in a 220-nm-thin silicon membrane perforated with a hexagonal pattern of holes (see also Fig. 1a of the main text). In the plane of the slab light is confined to this line by the photonic bandgap of the surrounding holes, and is confined to the silicon slab by total internal reflection. Our waveguide has a hole separation of $a$ = 420 nm and a hole radius of $r$ = 120 nm = $0.29a$.

The waveguide eigenmodes **E(r)** and **H(r)** were calculated with the MIT Photonic-Bands package [2]. We use a supercell of dimensions $a \times 11a\sqrt{3} \times 10h$, where $a$ is the lattice constant of the photonic crystal and $h$ is the thickness of the silicon slab. This supercell is sufficiently large to avoid interactions between neighboring supercells. The calculations were performed with a grid size of $a/16$, which ensures convergence of the eigenvalues to better than 0.1%. The refractive index of silicon used was modelled to be 3.48, which is suitable for wavelengths around 1570 nm.

## Supplementary Note 3: Number of unknowns

According to Eq. 1 (main text), in real space, the expected measurement for each probe position **R** can be calculated via the overlap integral of the experimental and reciprocal fields over the complete surface **S**. Here, the experimental fields at each position of the surface are unknown. Therefore, in real space, Eq. 1 contains an infinite number of unknowns for each, individual probe position (**R**$_{tip}$). In practice, the actual number of unknowns in this equation corresponds to 4*m*, where *m* is the number of **R**$_{tip}$ positions that we consider: basically, each component of the four in-plane fields in Eq. 1 ($\mathbf{E}^e_{x,y}$ and $\mathbf{H}^e_{x,y}$) at each position. In total, we only have 2*m* equations – one for $L_x$ and one for $L_y$ at each position – to deal with these 4*m* unknowns. Crucially, the equations are all interdependent since the signal at one point depends on the fields at all points, meaning that they all must be solved simultaneously. Computationally, this requires the inversion of a 2*m*X2*m* matrix.

Turning to reciprocal space (Eq. 3), we see that there are two equations and four unknowns for each plane wave. Although a reconstruction of the image requires solving Eq. 3 for all plane waves, the benefit of the k-space approach is that these equations are decoupled. That is, we only invert 2X2 matrices (Eq. 5), and we can gain insight to how the signal from each individual plane wave affects the measured signal. Importantly, as is the case with the real-space approach, we must still double the number of equations to match the number of unknowns, so that they can be solved. As we show in the main text, when working in reciprocal space, this can be done by specifying the direction in which the waves are propagating.

## Supplementary Note 4: The optical reciprocity theorem

### *The reciprocity integral*

We begin with the Maxwell equations

$$\nabla \times \mathbf{H}(\mathbf{r}) + i\omega \mathbf{D}(\mathbf{r}) = \mathbf{J}(\mathbf{r}),$$
$$\nabla \cdot \mathbf{D}(\mathbf{r}) = \rho(\mathbf{r}),$$
$$\nabla \times \mathbf{E}(\mathbf{r}) - i\omega \mathbf{B}(\mathbf{r}) = 0,$$
$$\nabla \cdot \mathbf{B}(\mathbf{r}) = 0,$$

(S1)

where all our fields are stationary,

$$f(\mathbf{r},t) = f(\mathbf{r})e^{-i\omega t} + c.c.,$$

and we consider a position and frequency dependent dielectric constant $\varepsilon(\mathbf{r}, \omega)$,

$$\mathbf{D}(\mathbf{r}) = \varepsilon_0 \varepsilon(\mathbf{r}, \omega) \mathbf{E}(\mathbf{r}),$$
$$\mathbf{B}(\mathbf{r}) = \mu_0 \mathbf{H}(\mathbf{r}).$$

(S2)

Note that for stationary fields the divergence conditions in (S1) follow immediately from the curl conditions and charge conservation.

We consider two solutions to equation (S1) corresponding to two sets of charge and current densities,

$$\nabla \times \mathbf{H}^1(\mathbf{r}) + i\omega \mathbf{D}^1(\mathbf{r}) = \mathbf{J}^1(\mathbf{r}),$$
$$\nabla \cdot \mathbf{D}^1(\mathbf{r}) = \rho^1(\mathbf{r}),$$
$$\nabla \times \mathbf{E}^1(\mathbf{r}) - i\omega \mathbf{B}^1(\mathbf{r}) = 0,$$
$$\nabla \cdot \mathbf{B}^1(\mathbf{r}) = 0,$$

(S3)

and

$$\nabla \times \mathbf{H}^2(\mathbf{r}) + i\omega \mathbf{D}^2(\mathbf{r}) = \mathbf{J}^2(\mathbf{r}),$$
$$\nabla \cdot \mathbf{D}^2(\mathbf{r}) = \rho^2(\mathbf{r}),$$
$$\nabla \times \mathbf{E}^2(\mathbf{r}) - i\omega \mathbf{B}^2(\mathbf{r}) = 0,$$
$$\nabla \cdot \mathbf{B}^2(\mathbf{r}) = 0.$$

(S4)

Then using

$$\nabla \cdot (\mathbf{X} \times \mathbf{Y}) = \mathbf{Y} \cdot (\nabla \times \mathbf{X}) - \mathbf{X} \cdot (\nabla \times \mathbf{Y}),$$

we have

$$\nabla \cdot (\mathbf{E}^1 \times \mathbf{H}^2) = \mathbf{H}^2 \cdot (\nabla \times \mathbf{E}^1) - \mathbf{E}^1 \cdot (\nabla \times \mathbf{H}^2)$$
$$= i\omega \mathbf{H}^2 \cdot \mathbf{B}^1 + i\omega \mathbf{E}^1 \cdot \mathbf{D}^2 - \mathbf{E}^1 \cdot \mathbf{J}^2,$$

and

$$\nabla \cdot \left( \mathbf{E}^2 \times \mathbf{H}^1 \right) = \mathbf{H}^2 \cdot \left( \nabla \times \mathbf{E}^2 \right) - \mathbf{E}^2 \cdot \left( \nabla \times \mathbf{H}^1 \right)$$
$$= i\omega \mathbf{H}^1 \cdot \mathbf{B}^2 + i\omega \mathbf{E}^2 \cdot \mathbf{D}^1 - \mathbf{E}^2 \cdot \mathbf{J}^1.$$

Since the two solutions (S3) and (S4) are at the same frequency we have

$$\mathbf{E}^1(\mathbf{r}) \cdot \mathbf{D}^2(\mathbf{r}) = \mathbf{E}^1(\mathbf{r}) \cdot \mathbf{E}^2(\mathbf{r}) \varepsilon_0 \varepsilon(\mathbf{r},\omega) = \mathbf{D}^1(\mathbf{r}) \cdot \mathbf{E}^2(\mathbf{r})$$

and of course

$$\mathbf{H}^2(\mathbf{r}) \cdot \mathbf{B}^1(\mathbf{r}) = \mathbf{H}^2(\mathbf{r}) \cdot \mathbf{H}^1(\mathbf{r}) \mu_0 = \mathbf{B}^2(\mathbf{r}) \cdot \mathbf{H}^1(\mathbf{r}),$$

so we find

$$\nabla \cdot \left( \mathbf{E}^1 \times \mathbf{H}^2 - \mathbf{E}^2 \times \mathbf{H}^1 \right) = -\mathbf{E}^1 \cdot \mathbf{J}^2 + \mathbf{E}^2 \cdot \mathbf{J}^1$$

or, moving to integral form,

$$\int_S \left( \mathbf{E}^1 \times \mathbf{H}^2 - \mathbf{E}^2 \times \mathbf{H}^1 \right) \cdot d\mathbf{s} = \int_V \left( -\mathbf{E}^1 \cdot \mathbf{J}^2 + \mathbf{E}^2 \cdot \mathbf{J}^1 \right) dv \qquad (S5)$$

*The scenario of interest*

We now look at our scenario of interest, shown in Fig. S2. We imagine some photonic-crystal structure – or more generally any sample – excited by a charge-current distribution $\left( \rho^1(\mathbf{r}), \mathbf{J}^1(\mathbf{r}) \right)$ located to the right of the structure and below the plane of interest $z_{plane}$, which lies in vacuum above the structure. We also imagine an observation structure – tip and fiber in our experiment – that is indicated by a cylinder in Fig. S2 and channels light from the neighborhood of $z_{plane}$ to the neighborhood of $z_{obs}$, the location of an observation plane. Its position in the $xy$-plane is specified by $\mathbf{R}_{tip} = (x_{tip}, y_{tip})$. In the plane of interest the electromagnetic field resulting from the driving source is $(\mathbf{E}^1(\mathbf{R}, \mathbf{R}_{tip}, z_{plane}), \mathbf{H}^1(\mathbf{R}, \mathbf{R}_{tip}, z_{plane}))$, where the dependence on $\mathbf{R}_{tip}$ reminds us that in general the field will be modified by the observation structure and its position. The electric field reaching the observation plane through the observation structure is $\mathbf{E}^1(\mathbf{R}_{tip}, z_{obs})$. This is the quantity we shall detect.

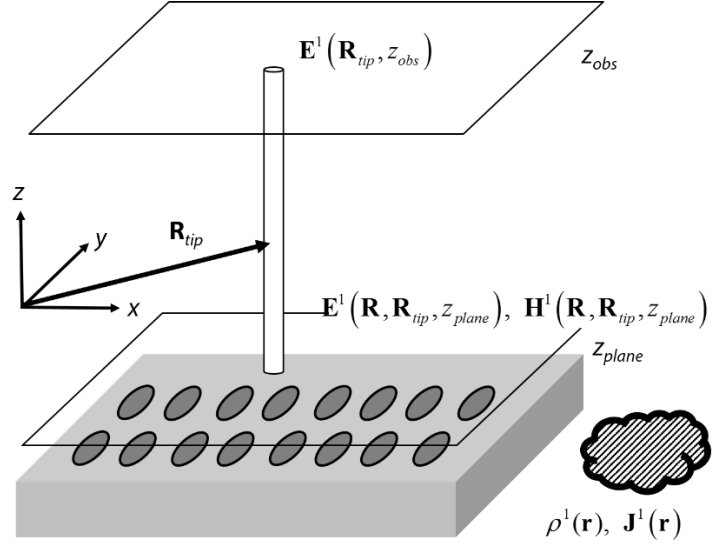

**Supplementary Figure S2: The scenario of interest.** The grey block and grey cloud indicate the photonic crystal (or more generally any sample) and charge current distribution, respectively. The observation structure (tip and fiber) are sketched as a vertical cylinder that is capped at the bottom and top by the observation and detection planes, respectively.

We describe the detection occurring by considering the overlap of this field with a second charge current distribution $(\rho^2(\mathbf{r}), \mathbf{J}^2(\mathbf{r}))$, which very generally we imagine in the neighborhood of $(\mathbf{R}_{tip}, z_{obs})$, see Fig. S3. The fields that this charge distribution generates in the plane of interest are denoted by $(\mathbf{E}^2(\mathbf{R}, \mathbf{R}_{tip}, z_{plane}), \mathbf{H}^2(\mathbf{R}, \mathbf{R}_{tip}, z_{plane}))$.

Now we construct a volume of interest as follows: Imagine a large circle in the plane $z_{plane}$, and cap it with a hemisphere. Then imagine increasing the radius of this circle to infinity, enlarging the hemisphere with it.

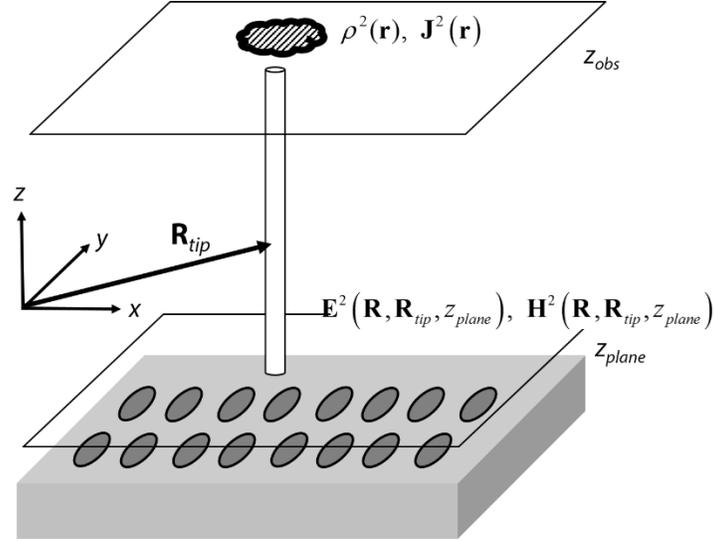

**Supplementary Figure S3: The reciprocal fields.** The grey block and grey cloud indicate the sample and reciprocal charge current distribution, respectively. The tip and fiber are sketched as a vertical cylinder, that is capped in the bottom and top by the observation and detection planes, respectively.

Consider first the surface integral appearing in (S5) over the hemisphere. The fields from both the first and the second charge-current distributions are responsible for generating the fields that appear, but as the radius of the hemisphere approaches infinity the Poynting-vector-like terms each drop off as $(radius)^{-2}$, and the difference will drop off faster. Thus only the integral over the plane at $z_{plane}$ (where $d\mathbf{s} = -\hat{\mathbf{z}}\, dxdy$) will contribute. With respect to the volume integral, only the second charge-current distribution ($\rho^2(\mathbf{r})$, $\mathbf{J}^2(\mathbf{r})$) lies in the volume, so only it will contribute. Thus (S5) will simplify here to

$$\int \left( \mathbf{E}^1(\mathbf{r}, \mathbf{R}_{tip}) \cdot \mathbf{J}^2(\mathbf{r}) \right) dv$$
$$= \int \left( \mathbf{E}^1(\mathbf{R}, \mathbf{R}_{tip}, z_{plane}) \times \mathbf{H}^2(\mathbf{R}, \mathbf{R}_{tip}, z_{plane}) - \mathbf{E}^2(\mathbf{R}, \mathbf{R}_{tip}, z_{plane}) \times \mathbf{H}^1(\mathbf{R}, \mathbf{R}_{tip}, z_{plane}) \right) \cdot \hat{\mathbf{z}}\, dxdy \quad \text{(S6)}$$

In practice we will employ a $\mathbf{J}^2(\mathbf{r})$ below that also depends on the position $\mathbf{R}_{tip}$ of the observation structure.

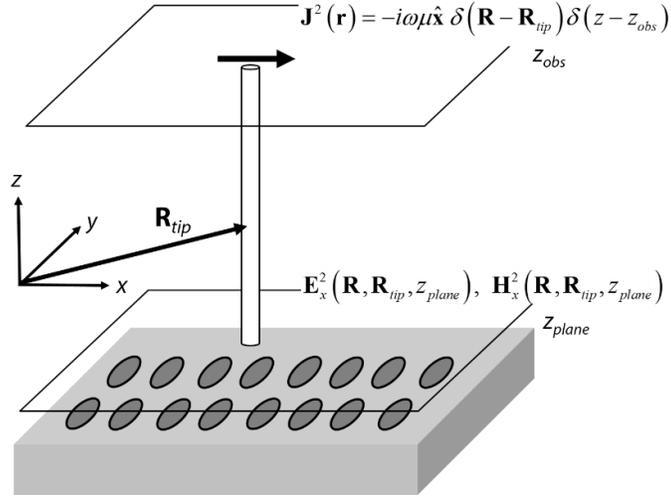

**Supplementary Figure S4: One of two dipole-moment directions considered.** The grey block indicates the Sample and the black arrow show the reciprocal dipole. The tip and fiber are sketched as a vertical cylinder, that is capped in the bottom and top by the observation and detection planes, respectively.

We now consider two special current densities $\mathbf{J}^2(\mathbf{r})$. Each corresponds to a point dipole $\boldsymbol{\mu}$ located at $(\mathbf{R}_{tip}, z_{obs})$; in one case we take the dipole oriented along the $\hat{\mathbf{x}}$ direction, and in the other we take it oriented along the $\hat{\mathbf{y}}$ direction. The first case is illustrated in Fig. 4; the current density is

$$\mathbf{J}^2(\mathbf{r}) = -i\omega\mu\hat{\mathbf{x}}\delta(\mathbf{R} - \mathbf{R}_{tip})\delta(z - z_{obs}).$$

For this orientation of the dipole, we denote the fields $(\mathbf{E}^2(\mathbf{R}, \mathbf{R}_{tip}, z_{plane}), \mathbf{H}^2(\mathbf{R}, \mathbf{R}_{tip}, z_{plane}))$ that result at $z = z_{plane}$ by $(\mathbf{E}_x^2(\mathbf{R}, \mathbf{R}_{tip}, z_{plane}), \mathbf{H}_x^2(\mathbf{R}, \mathbf{R}_{tip}, z_{plane}))$, and (S6) becomes

$$-i\omega\mu\left(\hat{\mathbf{x}} \cdot \mathbf{E}^1(\mathbf{R}_{tip}, z_{obs})\right)$$
$$= \int \left(\mathbf{E}^1(\mathbf{R}, \mathbf{R}_{tip}, z_{plane}) \times \mathbf{H}_x^2(\mathbf{R}, \mathbf{R}_{tip}, z_{plane}) - \mathbf{E}_x^2(\mathbf{R}, \mathbf{R}_{tip}, z_{plane}) \times \mathbf{H}^1(\mathbf{R}, \mathbf{R}_{tip}, z_{plane})\right) \cdot \hat{\mathbf{z}} \, dxdy,$$

and a similar expression can be written down if we take the dipole $\boldsymbol{\mu}$ to be oriented in the $\hat{\mathbf{y}}$ direction. If we put

$$L_i(\mathbf{R}_{tip}, z_{obs}) = -i\omega\mu\left(\hat{i} \cdot \mathbf{E}^1(\mathbf{R}_{tip}, z_{obs})\right),$$

where $i = x, y$, we then have

$$L_i(\mathbf{R}_{tip}, z_{obs})$$
$$= \int \left(\mathbf{E}^1(\mathbf{R}, \mathbf{R}_{tip}, z_{plane}) \times \mathbf{H}_i^2(\mathbf{R}, \mathbf{R}_{tip}, z_{plane}) - \mathbf{E}_i^2(\mathbf{R}, \mathbf{R}_{tip}, z_{plane}) \times \mathbf{H}^1(\mathbf{R}, \mathbf{R}_{tip}, z_{plane})\right) \cdot \hat{\mathbf{z}} \, dxdy \quad \text{(S7)}$$

So far everything is still exact. We assume that the quantities $L_i(\mathbf{R}_{tip}, z_{obs})$ are measurable; that is, both the amplitude and the phase of the complex quantities $(\hat{i} \cdot \mathbf{E}^1(\mathbf{R}_{tip}, z_{obs}))$ can be detected in the laboratory.

*The approximations*

We now introduce some approximations in the form of (S7). First, we assume that the fields $(\mathbf{E}^1(\mathbf{R}, \mathbf{R}_{tip}, z_{plane}), \mathbf{H}^1(\mathbf{R}, \mathbf{R}_{tip}, z_{plane}))$ are to good approximation unaffected by the presence of the observation structure. That is, at $z = z_{plane}$ we can calculate the fields $(\mathbf{E}^1(\mathbf{R}, \mathbf{R}_{tip}, z_{plane}), \mathbf{H}^1(\mathbf{R}, \mathbf{R}_{tip}, z_{plane}))$ as if the observation structure were not present (see Fig. S5).

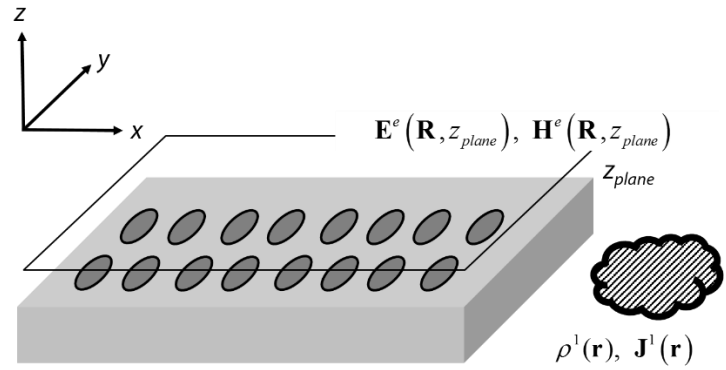

**Supplementary Figure S5: The approximation that the photonic-crystal fields at the plane of interest are unaffected by the observation structure.** The grey block and grey cloud indicate the sample and charge current distribution, respectively. The tip is not sketched, to illustrate that we assume that it does not affect the experimental fields.

Within this approximation the dependence of $(\mathbf{E}^1(\mathbf{R}, \mathbf{R}_{tip}, z_{plane}), \mathbf{H}^1(\mathbf{R}, \mathbf{R}_{tip}, z_{plane}))$ on $\mathbf{R}_{tip}$ vanishes, and we put $(\mathbf{E}^1(\mathbf{R}, \mathbf{R}_{tip}, z_{plane}), \mathbf{H}^1(\mathbf{R}, \mathbf{R}_{tip}, z_{plane})) \rightarrow (\mathbf{E}^e(\mathbf{R}, z_{plane}), \mathbf{H}^e(\mathbf{R}, z_{plane}))$ in (S7), where the superscript $e$ indicates the fields in the experimental situation discussed in the main text. Second, we assume that the fields $(\mathbf{E}_i^2(\mathbf{R}, \mathbf{R}_{tip}, z_{plane}), \mathbf{H}_i^2(\mathbf{R}, \mathbf{R}_{tip}, z_{plane}))$ are to good approximation unaffected by the presence of the sample. Thus, they can be evaluated as if the photonic crystal were not present; see Fig. S6.

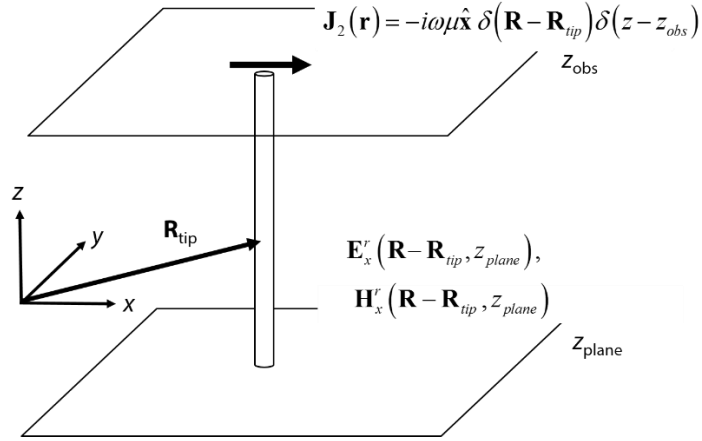

**Supplementary Figure S6: The approximation that the reciprocal fields at the plane of interest are unaffected by the sample.** The black arrow show the dipole current that sets up the reciprocal fields, which we assume are unperturbed by the sample.

Then, although these fields at $z = z_{plane}$ do still depend on $\mathbf{R}_{tip}$, they will only depend on the difference $\mathbf{R} - \mathbf{R}_{tip}$, and we can write

$$(\mathbf{E}_i^2(\mathbf{R},\mathbf{R}_{tip},z_{plane}),\ \mathbf{H}_i^2(\mathbf{R},\mathbf{R}_{tip},z_{plane})) \to (\mathbf{E}_i^r(\mathbf{R}-\mathbf{R}_{tip},z_{plane}),\ \mathbf{H}_i^r(\mathbf{R}-\mathbf{R}_{tip},z_{plane})),$$

where the superscript $r$ indicates the fields in the reciprocal situation discussed in the main text. Using these approximations in (S7) we have

$$L_i(\mathbf{R}_{tip}, z_{obs})$$
$$= \int_S \left( \mathbf{E}^e(\mathbf{R},z_{plane}) \times \mathbf{H}_i^r(\mathbf{R}-\mathbf{R}_{tip},z_{plane}) - \mathbf{E}_i^r(\mathbf{R}-\mathbf{R}_{tip},z_{plane}) \times \mathbf{H}^e(\mathbf{R},z_{plane}) \right) \cdot \hat{\mathbf{z}}\ dxdy. \quad (S8)$$

Please note that, for clarity and brevity, we have omitted the dependence on $z_{obs}$ and $z_{plane}$ in the main text. The former affects only the overall amplitude scaling of the signals. Throughout our manuscript we omit this dependence and only compare the relative strengths between field components and signals. The latter, $z_{plane}$, is fixed to 10 nm below the probe apex throughout this work.

## Supplementary Note 5: Basis Conversion

### *The move to $\kappa$ space*

Next we Fourier decompose in the *xy*-plane, writing

$$\mathbf{E}^e(\mathbf{R}, z) = \int \frac{d\boldsymbol{\kappa}}{(2\pi)^2} \mathbf{E}^e(\boldsymbol{\kappa}, z) e^{i\boldsymbol{\kappa}\cdot\mathbf{R}},$$

etc., where $d\boldsymbol{\kappa} = d\kappa_x d\kappa_y$. Then, we can write

$$L_i(\mathbf{R}_{tip}, z_{obs}) = \int_S \int \frac{d\boldsymbol{\kappa} d\boldsymbol{\kappa}'}{(2\pi)^4} \mathcal{F}(\boldsymbol{\kappa}, \boldsymbol{\kappa}') \, e^{i\boldsymbol{\kappa}\cdot\mathbf{R}} e^{i\boldsymbol{\kappa}'\cdot(\mathbf{R}-\mathbf{R}_{tip})} dxdy,$$

where

$$\mathcal{F}(\boldsymbol{\kappa}, \boldsymbol{\kappa}') = \left( \mathbf{E}^e(\boldsymbol{\kappa}, z_{plane}) \times \mathbf{H}_i^r(\boldsymbol{\kappa}', z_{plane}) - \mathbf{E}_i^r(\boldsymbol{\kappa}', z_{plane}) \times \mathbf{H}^e(\boldsymbol{\kappa}, z_{plane}) \right) \cdot \hat{\mathbf{z}}$$

The integral over $\mathbf{R} = (x, y)$ vanishes unless $\boldsymbol{\kappa}' = -\boldsymbol{\kappa}$, so from (S8) we have

$$L_i(\mathbf{R}_{tip}, z_{obs})$$
$$= \int \frac{d\boldsymbol{\kappa}}{(2\pi)^2} \left( \mathbf{E}^e(\boldsymbol{\kappa}, z_{plane}) \times \mathbf{H}_i^r(-\boldsymbol{\kappa}, z_{plane}) - \mathbf{E}_i^r(-\boldsymbol{\kappa}, z_{plane}) \times \mathbf{H}^e(\boldsymbol{\kappa}, z_{plane}) \right) \cdot \hat{\mathbf{z}} \, e^{i\boldsymbol{\kappa}\cdot\mathbf{R}_{tip}}.$$

Now putting

$$L_i(\mathbf{R}_{tip}, z_{obs}) = \int \frac{d\boldsymbol{\kappa}}{(2\pi)^2} L_i(\boldsymbol{\kappa}, z_{obs}) e^{i\boldsymbol{\kappa}\cdot\mathbf{R}_{tip}},$$

we see that

$$L_i(\boldsymbol{\kappa}, z_{obs}) = \left( \mathbf{E}^e(\boldsymbol{\kappa}, z_{plane}) \times \mathbf{H}_i^r(-\boldsymbol{\kappa}, z_{plane}) - \mathbf{E}_i^r(-\boldsymbol{\kappa}, z_{plane}) \times \mathbf{H}^e(\boldsymbol{\kappa}, z_{plane}) \right) \cdot \hat{\mathbf{z}}$$

or, writing out the components,

$$L_i(\boldsymbol{\kappa}, z_{obs}) = E_x^e(\boldsymbol{\kappa}, z_{plane}) H_{i,y}^r(-\boldsymbol{\kappa}, z_{plane}) - E_y^e(\boldsymbol{\kappa}, z_{plane}) H_{i,x}^r(-\boldsymbol{\kappa}, z_{plane})$$
$$- E_{i,x}^r(-\boldsymbol{\kappa}, z_{plane}) H_y^e(\boldsymbol{\kappa}, z_{plane}) + E_{i,y}^r(-\boldsymbol{\kappa}, z_{plane}) H_x^e(\boldsymbol{\kappa}, z_{plane}). \quad \text{(S9)}$$

### *Fields in vacuum*

Now let us consider the nature of an electromagnetic in the neighborhood of $z$ where, within that neighborhood, there are no sources of any sort. Then in the neighborhood of $z$ the Maxwell equations (S1) reduce to

$$\nabla \times \mathbf{H} + i\omega \mathbf{D} = 0,$$
$$\nabla \cdot \mathbf{E} = 0,$$
$$\nabla \times \mathbf{E} - i\omega \mathbf{B} = 0,$$
$$\nabla \cdot \mathbf{B} = 0,$$

with $\mathbf{D} = \epsilon_0 \mathbf{E}$, $\mathbf{B} = \mu_0 \mathbf{H}$. Fourier transforming the fields in the *xy*-plane,

$$\mathbf{E}(\mathbf{r}) = \mathbf{E}(\mathbf{R}, z) = \int \frac{d\kappa}{(2\pi)^2} \mathbf{E}(\kappa, z) e^{i\kappa \cdot \mathbf{R}},$$
$$\mathbf{H}(\mathbf{r}) = \mathbf{H}(\mathbf{R}, z) = \int \frac{d\kappa}{(2\pi)^2} \mathbf{H}(\kappa, z) e^{i\kappa \cdot \mathbf{R}},$$
(S10)

the terms $\mathbf{E}(\kappa, z)$ and $\mathbf{H}(\kappa, z)$ are composed of upward propagating (or evanescent) waves and downward propagating (or evanescent) waves. That is, we have

$$\mathbf{E}(\kappa, z) = \left(\hat{\mathbf{s}} E_{s+}(\kappa) + \hat{\mathbf{p}}_+ E_{p+}(\kappa)\right) e^{ik_z z}$$
$$+ \left(\hat{\mathbf{s}} E_{s-}(\kappa) + \hat{\mathbf{p}}_- E_{p-}(\kappa)\right) e^{-ik_z z}$$
(S11)

where

$$k_z = \sqrt{k_0^2 - \kappa^2},$$

with $k_0 = \omega/c$; since here the argument of the square root is always real, we take $k_z$ to be either a positive real number or a positive imaginary number. This guarantees that the + fields are associated with upward propagating (or evanescent) fields and the – fields are associated with downward propagating (or evanescent) fields. And here

$$\hat{\mathbf{s}} = \hat{\mathbf{\kappa}} \times \hat{\mathbf{z}},$$
(S12)

and

$$\hat{\mathbf{p}}_\pm = \frac{\kappa \hat{\mathbf{z}} \mp k_z \hat{\mathbf{\kappa}}}{k_0}.$$
(S13)

For each $\kappa$ there are four independent quantities in (S11), $E_{s+}(\kappa)$, $E_{p+}(\kappa)$, $E_{s-}(\kappa)$, and $E_{p-}(\kappa)$. Since from the Maxwell equations we can write the corresponding expression for $\mathbf{H}(\kappa, z)$ as

$$\mathbf{H}(\kappa, z) = \frac{\left(\hat{\mathbf{s}} E_{p+}(\kappa) - \hat{\mathbf{p}}_+ E_{s+}(\kappa)\right)}{Z_0} e^{ik_z z}$$
$$+ \frac{\left(\hat{\mathbf{s}} E_{p-}(\kappa) - \hat{\mathbf{p}}_- E_{s-}(\kappa)\right)}{Z_0} e^{-ik_z z}$$
(S14)

there are no additional independent quantities; we have only upward propagating (or evanescent) waves of $s$- and $p$-polarization type, and downward propagating (or evanescent) waves of $s$- and $p$-polarization type.

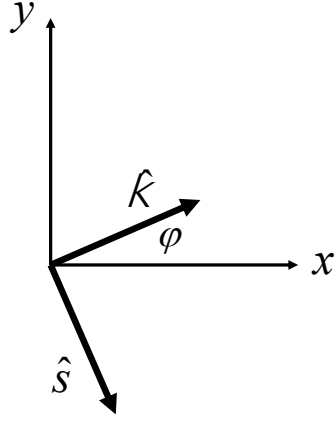

**Supplementary Figure S7: Unit vectors**

## *The experimental fields*

Because we measure above the sample, we can assume for the experimental fields that we have no amplitudes propagating (or evanescent) in the downward direction. Then from (S11) and (S14) we have

$$\mathbf{E}^e(\boldsymbol{\kappa}, z) = \left(\hat{\mathbf{s}} E_{s+}(\boldsymbol{\kappa}) + \hat{\mathbf{p}}_+ E_{p+}(\boldsymbol{\kappa})\right) e^{ik_z z},$$

$$\mathbf{H}^e(\boldsymbol{\kappa}, z) = \frac{\left(\hat{\mathbf{s}} E_{p+}(\boldsymbol{\kappa}) - \hat{\mathbf{p}}_+ E_{s+}(\boldsymbol{\kappa})\right)}{Z_0} e^{ik_z z}, \qquad (S15)$$

and so

$$\mathbf{E}^e(\boldsymbol{\kappa}, z_{plane}) = \hat{\mathbf{s}} E_{s+}(\boldsymbol{\kappa}, z_{plane}) + \hat{\mathbf{p}}_+ E_{p+}(\boldsymbol{\kappa}, z_{plane}),$$

$$\mathbf{H}^e(\boldsymbol{\kappa}, z_{plane}) = \hat{\mathbf{s}} \frac{E_{p+}(\boldsymbol{\kappa}, z_{plane})}{Z_0} - \hat{\mathbf{p}}_+ \frac{E_{s+}(\boldsymbol{\kappa}, z_{plane})}{Z_0}, \qquad (S16)$$

where

$$E_{s,p+}(\boldsymbol{\kappa}, z) \equiv E_{s,p+}(\boldsymbol{\kappa}) e^{ik_z z}. \qquad (S17)$$

Then using (S13) we have

$$E_x^e(\boldsymbol{\kappa}, z_{plane}) = (\hat{\mathbf{x}} \cdot \hat{\mathbf{s}}) E_{s+}(\boldsymbol{\kappa}, z_{plane}) - \frac{k_z}{k_0}(\hat{\mathbf{x}} \cdot \hat{\boldsymbol{\kappa}}) E_{p+}(\boldsymbol{\kappa}, z_{plane}),$$

$$E_y^e(\boldsymbol{\kappa}, z_{plane}) = (\hat{\mathbf{y}} \cdot \hat{\mathbf{s}}) E_{s+}(\boldsymbol{\kappa}, z_{plane}) - \frac{k_z}{k_0}(\hat{\mathbf{y}} \cdot \hat{\boldsymbol{\kappa}}) E_{p+}(\boldsymbol{\kappa}, z_{plane}),$$

$$H_x^e(\boldsymbol{\kappa}, z_{plane}) = (\hat{\mathbf{x}} \cdot \hat{\mathbf{s}}) \frac{E_{p+}(\boldsymbol{\kappa}, z_{plane})}{Z_0} + \frac{k_z}{k_0}(\hat{\mathbf{x}} \cdot \hat{\boldsymbol{\kappa}}) \frac{E_{s+}(\boldsymbol{\kappa}, z_{plane})}{Z_0},$$

$$H_y^e(\boldsymbol{\kappa}, z_{plane}) = (\hat{\mathbf{y}} \cdot \hat{\mathbf{s}}) \frac{E_{p+}(\boldsymbol{\kappa}, z_{plane})}{Z_0} + \frac{k_z}{k_0}(\hat{\mathbf{y}} \cdot \hat{\boldsymbol{\kappa}}) \frac{E_{s+}(\boldsymbol{\kappa}, z_{plane})}{Z_0}.$$

We introduce an angle $\phi$ which indicates the direction that $\hat{\boldsymbol{\kappa}}$ makes from the $\hat{\mathbf{x}}$ axis in the $xy$ plane,

$$\begin{aligned}\hat{\boldsymbol{\kappa}} &= \hat{\mathbf{x}}\cos\phi + \hat{\mathbf{y}}\sin\phi, \\ \hat{\mathbf{s}} &= \hat{\mathbf{x}}\sin\phi - \hat{\mathbf{y}}\cos\phi,\end{aligned} \quad (S18)$$

see Fig. S7, in terms of which we have

$$\begin{aligned}E_x^e(\boldsymbol{\kappa}, z_{plane}) &= \sin\phi\, E_{s+}(\boldsymbol{\kappa}, z_{plane}) - \frac{k_z}{k_0}\cos\phi\, E_{p+}(\boldsymbol{\kappa}, z_{plane}), \\ E_y^e(\boldsymbol{\kappa}, z_{plane}) &= -\cos\phi\, E_{s+}(\boldsymbol{\kappa}, z_{plane}) - \frac{k_z}{k_0}\sin\phi\, E_{p+}(\boldsymbol{\kappa}, z_{plane}), \\ H_x^e(\boldsymbol{\kappa}, z_{plane}) &= \sin\phi\, \frac{E_{p+}(\boldsymbol{\kappa}, z_{plane})}{Z_0} + \frac{k_z}{k_0}\cos\phi\, \frac{E_{s+}(\boldsymbol{\kappa}, z_{plane})}{Z_0}, \\ H_y^e(\boldsymbol{\kappa}, z_{plane}) &= -\cos\phi\, \frac{E_{p+}(\boldsymbol{\kappa}, z_{plane})}{Z_0} + \frac{k_z}{k_0}\sin\phi\, \frac{E_{s+}(\boldsymbol{\kappa}, z_{plane})}{Z_0}.\end{aligned} \quad (S19)$$

From **Error! Reference source not found.** we then have

$$\begin{aligned}L_i(\boldsymbol{\kappa}, z_{obs}) = &\left(\sin\phi\, E_{s+}(\boldsymbol{\kappa}, z_{plane}) - \frac{k_z}{k_0}\cos\phi\, E_{p+}(\boldsymbol{\kappa}, z_{plane})\right) H_{i,y}^r(-\boldsymbol{\kappa}, z_{plane}) \\ &-\left(-\cos\phi\, E_{s+}(\boldsymbol{\kappa}, z_{plane}) - \frac{k_z}{k_0}\sin\phi\, E_{p+}(\boldsymbol{\kappa}, z_{plane})\right) H_{i,x}^r(-\boldsymbol{\kappa}, z_{plane}) \\ &- E_{i,x}^r(-\boldsymbol{\kappa}, z_{plane})\left(-\cos\phi\, \frac{E_{p+}(\boldsymbol{\kappa}, z_{plane})}{Z_0} + \frac{k_z}{k_0}\sin\phi\, \frac{E_{s+}(\boldsymbol{\kappa}, z_{plane})}{Z_0}\right) \\ &+ E_{i,y}^r(-\boldsymbol{\kappa}, z_{plane})\left(\sin\phi\, \frac{E_{p+}(\boldsymbol{\kappa}, z_{plane})}{Z_0} + \frac{k_z}{k_0}\cos\phi\, \frac{E_{s+}(\boldsymbol{\kappa}, z_{plane})}{Z_0}\right),\end{aligned}$$

or

$$Z_0 L_i(\boldsymbol{\kappa}, z_{obs}) = N_{i,s}(\boldsymbol{\kappa}) E_{s+}(\boldsymbol{\kappa}, z_{plane}) + N_{i,p}(\boldsymbol{\kappa}) E_{p+}(\boldsymbol{\kappa}, z_{plane}), \quad (S20)$$

where

$$N_{i,s}(\boldsymbol{\kappa}) = \sin\phi\, Z_0 H^r_{i,y}(-\boldsymbol{\kappa}, z_{plane}) + \cos\phi\, Z_0 H^r_{i,x}(-\boldsymbol{\kappa}, z_{plane})$$
$$-\frac{k_z}{k_0}\sin\phi\, E^r_{i,x}(-\boldsymbol{\kappa}, z_{plane}) + \frac{k_z}{k_0}\cos\phi\, E^r_{i,y}(-\boldsymbol{\kappa}, z_{plane}),$$

and

$$N_{i,p}(\boldsymbol{\kappa}) = -\frac{k_z}{k_0}\cos\phi\, Z_0 H^r_{i,y}(-\boldsymbol{\kappa}, z_{plane}) + \frac{k_z}{k_0}\sin\phi\, Z_0 H^r_{i,x}(-\boldsymbol{\kappa}, z_{plane})$$
$$+\cos\phi\, E^r_{i,x}(-\boldsymbol{\kappa}, z_{plane}) + \sin\phi\, E^r_{i,y}(-\boldsymbol{\kappa}, z_{plane}).$$

We can write (S20) as

$$Z_0 \begin{bmatrix} L_x(\boldsymbol{\kappa}, z_{obs}) \\ L_y(\boldsymbol{\kappa}, z_{obs}) \end{bmatrix} = \mathrm{N}(\boldsymbol{\kappa}) \begin{bmatrix} E_{s+}(\boldsymbol{\kappa}, z_{plane}) \\ E_{p+}(\boldsymbol{\kappa}, z_{plane}) \end{bmatrix}, \tag{S21}$$

where for each $\boldsymbol{\kappa}$ the $2\times 2$ matrix $\mathrm{N}(\boldsymbol{\kappa})$ is given by

$$\mathrm{N}(\boldsymbol{\kappa}) = \begin{bmatrix} N_{x,s}(\boldsymbol{\kappa}) & N_{x,p}(\boldsymbol{\kappa}) \\ N_{y,s}(\boldsymbol{\kappa}) & N_{y,p}(\boldsymbol{\kappa}) \end{bmatrix},$$

and as long as $\det \mathrm{N}(\boldsymbol{\kappa}) \neq 0$ we can invert (S21) to give

$$\begin{bmatrix} E_{s+}(\boldsymbol{\kappa}, z_{plane}) \\ E_{p+}(\boldsymbol{\kappa}, z_{plane}) \end{bmatrix} = Z_0 \mathrm{N}^{-1}(\boldsymbol{\kappa}) \begin{bmatrix} L_x(\boldsymbol{\kappa}, z_{obs}) \\ L_y(\boldsymbol{\kappa}, z_{obs}) \end{bmatrix}$$

Once this is determined we can find the electric field and the magnetic field anywhere above the sample by using (S15) and (S17),

$$\mathbf{E}^e(\boldsymbol{\kappa}, z) = \left(\hat{\mathbf{s}} E_{s+}(\boldsymbol{\kappa}, z_{plane}) + \hat{\mathbf{p}}_+ E_{p+}(\boldsymbol{\kappa}, z_{plane})\right) e^{ik_z(z-z_{plane})},$$
$$Z_0 \mathbf{H}^e(\boldsymbol{\kappa}, z) = \left(\hat{\mathbf{s}} E_{p+}(\boldsymbol{\kappa}, z_{plane}) - \hat{\mathbf{p}}_+ E_{s+}(\boldsymbol{\kappa}, z_{plane})\right) e^{ik_z(z-z_{plane})},$$

and so, using (S10),

$$\mathbf{E}^e(\mathbf{r}) = \int \frac{d\boldsymbol{\kappa}}{(2\pi)^2} \left(\hat{\mathbf{s}} E_{s+}(\boldsymbol{\kappa}, z_{plane}) + \hat{\mathbf{p}}_+ E_{p+}(\boldsymbol{\kappa}, z_{plane})\right) e^{i\boldsymbol{\kappa}\cdot\mathbf{R}} e^{ik_z(z-z_{plane})},$$
$$Z_0 \mathbf{H}^e(\mathbf{r}) = \int \frac{d\boldsymbol{\kappa}}{(2\pi)^2} \left(\hat{\mathbf{s}} E_{p+}(\boldsymbol{\kappa}, z_{plane}) - \hat{\mathbf{p}}_+ E_{s+}(\boldsymbol{\kappa}, z_{plane})\right) e^{i\boldsymbol{\kappa}\cdot\mathbf{R}} e^{ik_z(z-z_{plane})}.$$

the Cartesian components can be extracted. Again, please note that for clarity and brevity we have omitted the dependence on $z_{obs}$ and $z_{plane}$ in the main text.

## Supplementary Note 6: Plasmonic nanowire field retrieval

### *Fabricating the plasmonic nanowires*

To couple light to the nanowire we use a metal hole-array and waveguide-taper [3]. The nanowire, waveguide-taper, and hole-array are patterned by electron beam lithography into a bilayer PMMA resist. The Au is evaporated through resistive heating on the patterned sample, followed by liftoff. The hole-array has a pitch of 1 µm and a hole-diameter of roughly 0.5 µm. The nanowire length is approximately 50 µm.

### *Plasmonic nanowire mode calculations*

The optical modes in the 130 nm wide and 50 nm thick plasmonic nanowire were calculated with a COMSOL 2D-eigenmode analysis. We use a wavelength of 1550 nm, a refractive index of 1.5 for the BK7, and Johnson and Christy values for the gold [4]. Edges of the nanowire were rounded with a 20nm radius of curvature. We extract the fields from these calculations along the path shown in Fig. S8.

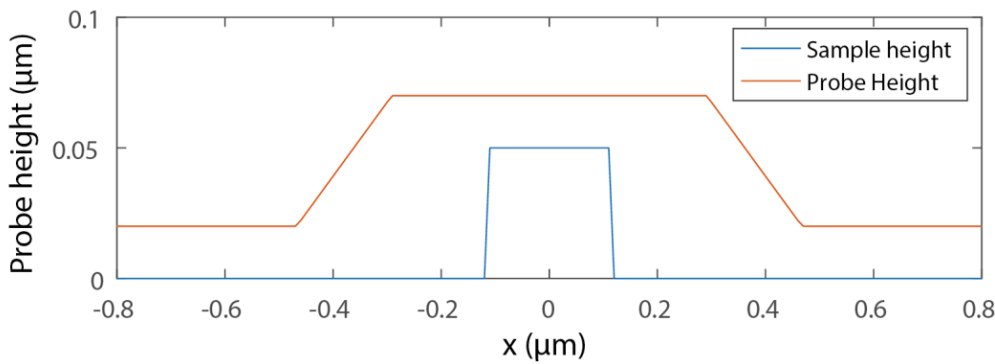

**Supplementary Figure S8: Nanowire scan path.** The blue line indicates the top of the sample in a cross section along *x*. We extracted the data from the simulations of the plasmonic nanowire along the red line.

### *Polarization mixing removal*

For the nanowire studies we use the measurements presented in [5]. Those measurements contain a small degree of polarization mixing. Such polarization mixing can arise if the waveplates of the setup are not set completely perfectly, if the tip has a slight asymmetry. Light that is emitted $x$-polarized from the tip, is detected not only on $L_x$ but also on $L_y$ and similarly for light emitted $y$-polarized. Fortunately, we can identify and filter this mixing by using the symmetry properties of the structures' optical fields. The optical fields of the mode of a nanophotonic structure match the structures' symmetries [6]. However, the measured fields above the plasmonic nanowire are not symmetric about the center of the waveguide. That is, if we were to assume the center of the waveguide was at $x=0$, $L_y$ would be nearly symmetrical, but $L_x$ would not.

Such a breaking of symmetry is indicative of polarization mixing [6]. Here, we employ the approach we presented earlier in Ref. [6] to remove this mixing. The essence of this removal lies in the symmetries of the fields of a TM mode. For a TM mode, $E_y$, $E_z$, and $H_x$ have an even symmetry around the waveguide center, whereas the other components have odd symmetry. As a result, $L_x$ has even symmetry about $x=0$, whereas that in $L_y$ has odd symmetry. By mirroring both maps around $x=0$ and adding or subtracting the mirrored to the original maps, we can obtain the even and odd symmetry contributions to both channels.

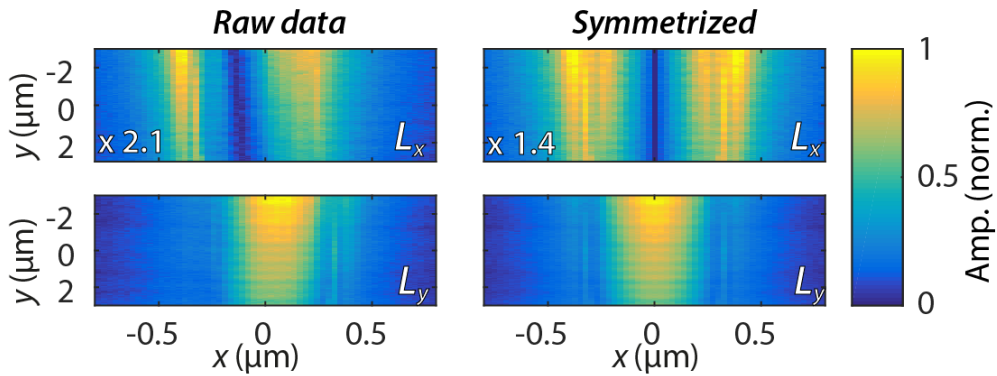

**Supplementary Figure S9: Measured field maps on a plasmonic nanowire.** The left column shows the amplitude of the signal measured on $L_x$ and $L_y$. The right column show the odd symmetry (top panel) and even symmetry (bottom panel) contributions to $L_x$ and $L_y$. The color of all maps is scaled to the maximum amplitude (norm.); scaling of $L_x$ relative to $L_y$ is indicated by the multiplication factors in the bottom left of the top panels.

Supplementary Figure S9 shows the results of this approach. The two panels in the left column show the raw measured amplitudes. We symmetrize these panels around the $x=0$ center of the waveguide. The symmetrized signals are shown in the right panels. As explained in the main text we retrieve the experimental optical fields from these symmetrized fields.

*Field retrieval*

We now insert these symmetrized fields in the deconvolution algorithm to retrieve the maps of the experimental optical fields. These field maps, which are shown in Fig. S10, qualitatively agree with the calculated maps. That is, $E_x$, and $H_y$ (and $E_y$, $H_x$) show a zero (and a maximum) in the center, reminiscent of an odd (and even) symmetry fields. Further, the $H_y$ component we retrieve is more spread out in $x$ than $E_x$, which we also see in our calculated fields. Likewise, in both experiment and theory $H_x$, is more confined than $E_y$. Notably, the agreement between experiment and theory is further established by the enhanced side lobes that are visible in both the calculated and retrieved $H_x$ but not in $E_y$. We also observe that all retrieved fields are slightly less confined than the calculated fields, which we attribute to the finite size of the Fourier filter we used.

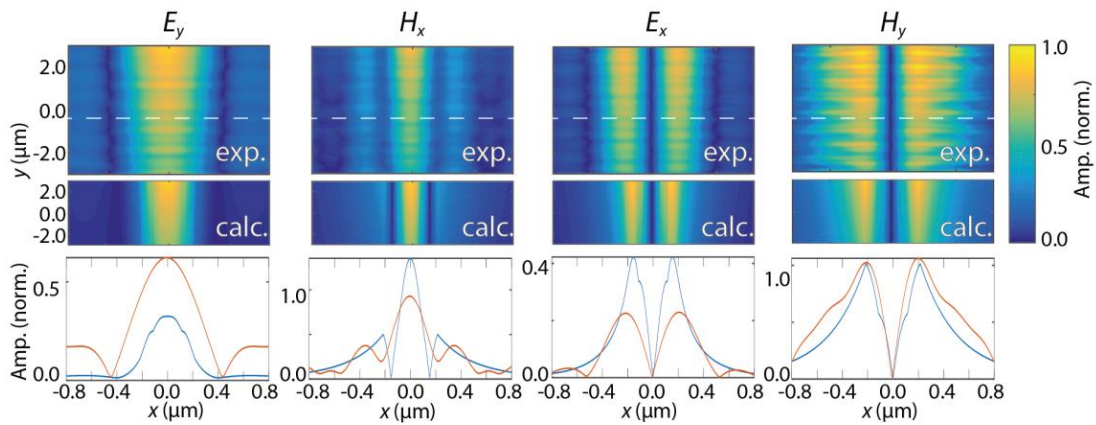

**Supplementary Figure S10: Retrieved nanowire electric and magnetic fields.** Panels show two-dimensional maps and line cuts of the calculated and reconstructed electric and magnetic fields above the plasmonic nanowire. The top (and middle) row of panels show the amplitude of the retrieved and calculated field maps, respectively. Each panel is scaled to its maximum. The bottom row of panels shows line cuts taken along the white dashed lines in the field maps. Red and blue lines correspond to line cuts through the fields reconstructed from the experimental data and calculated fields, respectively.

## Supplementary Note 7: Setting the correct maximum filter

To select a value for $\kappa_{max}$ we follow an empirical approach. That is, for $\kappa_{max} = 0$ to $\kappa_{max} = 30k_0$, we calculate the mean (of the absolute of the) difference in the electric field amplitude from pixel to pixel in the retrieved distributions. Fig. S11, which depicts the results of this approach, clearly shows a low amount of pixel to pixel noise for $\kappa_{max} \leq 9k_0$ (the first plateau), after which this difference rapidly increases by an order of magnitude. Furthermore, when $\kappa_{max} \leq 2k_0$ the pixel to pixel noise falls off, but those filters are practically not relevant because they will miss essential features of the near field. Combined, these observations suggest that filter $\kappa_{max}$ values on the first plateau are most suitable.

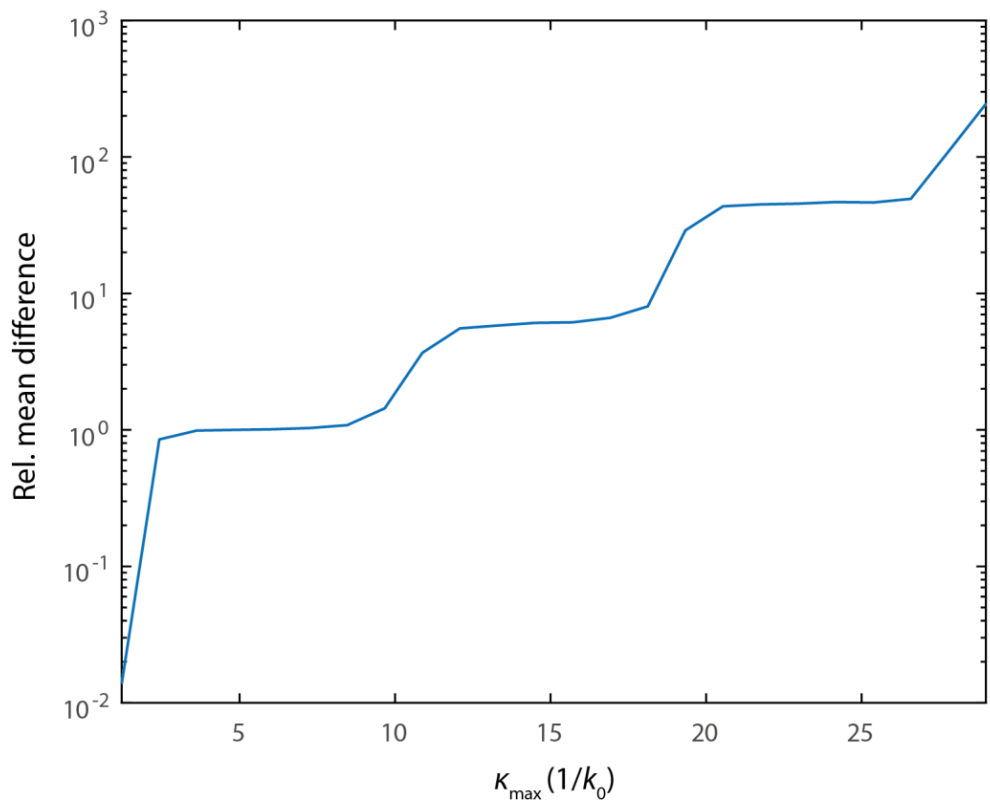

**Supplementary Figure S11: Mean difference dependence of filter value.** The mean is taken over all fields components and the y-axis is normalized to the difference when $\kappa_{max} = 2k_0$, and is shown on a logarithmic scale. This figure was generated at a mean noise level of 0.2 of the maximum signal amplitude.

This observation is supported by Fig. S12, which shows the retrieved fields for $\kappa_{max}$ values directly before, on and beyond that first plateau. Evidently, $\kappa_{max}$ values before the plateau miss essential features in the retrieved fields. Likewise, values larger than the $\kappa_{max}$ corresponding to the first plateau ($\kappa_{max} \geq 12k_0$) rapidly increase the noise in the retrieved fields, and clearly do not yield physically meaningful results. This empirical approach allows us to conclude that our choice of $\kappa_{max} = 5k_0$ is correct (note that there is no qualitative difference between $\kappa_{max} = 3k_0$ to $\kappa_{max} = 9k_0$).

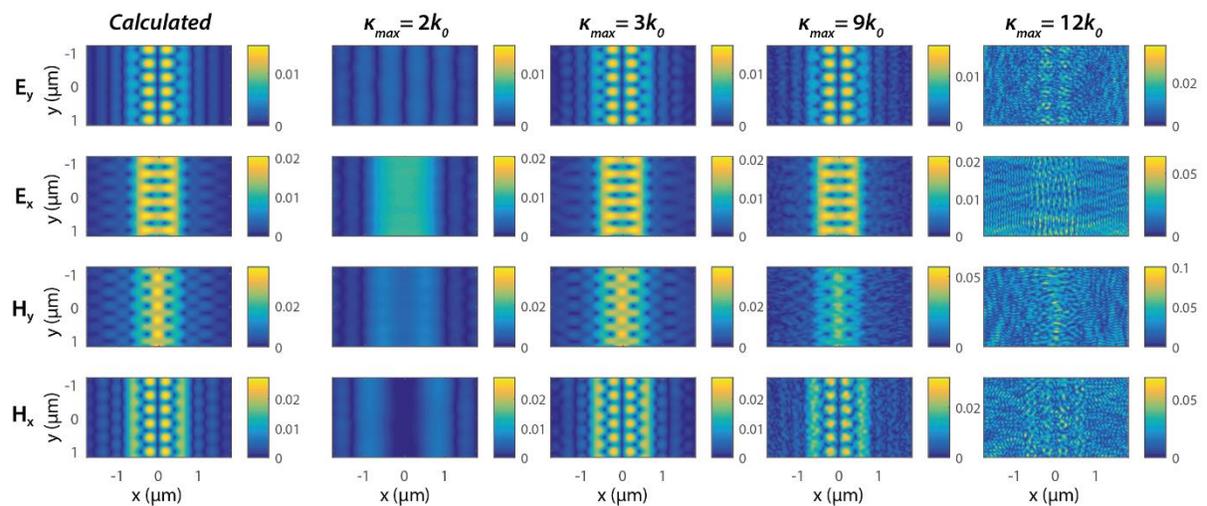

**Supplementary Figure S12: Effect of filtering on retrieved fields.** Panels show calculated retrieved electric and magnetic field maps for various filters, as indicated above each column. These panels were generated at a mean noise level of 0.2 of the maximum signal amplitude. Panel rows show the different electric and magnetic field components. The color bars next to each panel show the normalized field amplitude.

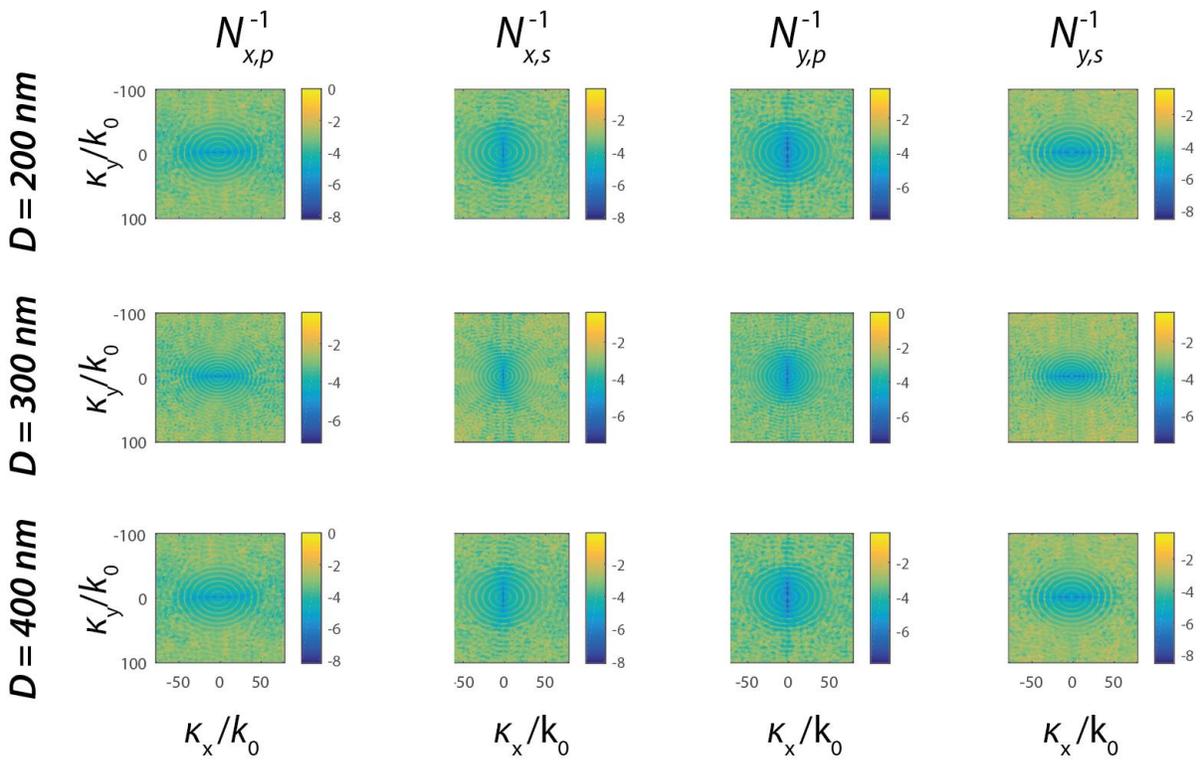

**Supplementary Figure S13: Effect of probe size on deconvolution matrix.** Each row of panels shows all components of $\log_{10}(N^{-1})$ for the probe diameter ($D$) written next to that row. The component that each column represents is indicated above that column.

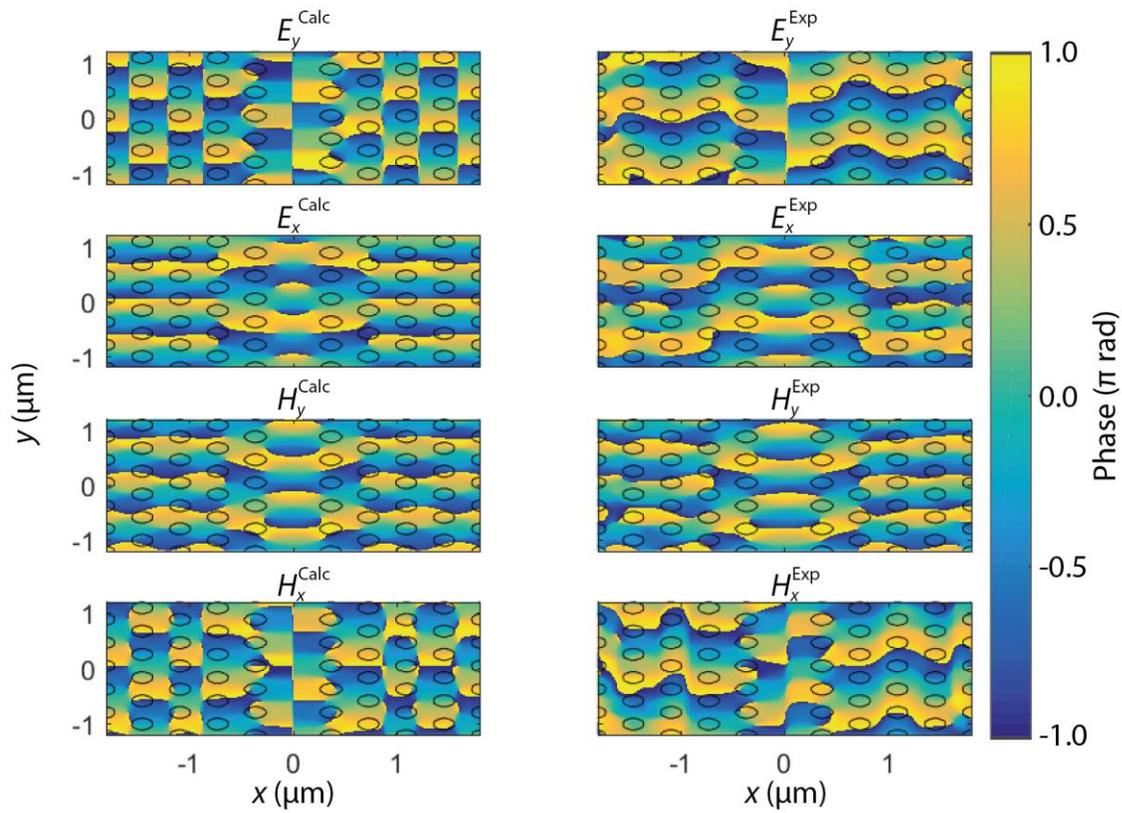

**Supplementary Figure S14: Retrieved PhCW phase maps.** The left (and right) column of panels show the calculated (and retrieved) phase maps of the electric and magnetic fields 280 nm above the PhCW. Each row of panels shows the component of the field indicated above the phase maps in that row.